# Electronic Transport through DNA Nucleotides in a BC$_3$ Nanogap for Rapid DNA Sequencing

Rameshwar L. Kumawat[†], Priyanka Garg[#], Gargee Bhattacharyya[†], Biswarup Pathak*,[†,#],

[†]Discipline of Metallurgy Engineering and Materials Science, [#]Discipline of Chemistry, School of Basic Sciences, Indian Institute of Technology (IIT) Indore, Indore, Madhya Pradesh, 453552, India

*E-mail: biswarup@iiti.ac.in

**Abstract**

Recently solid-state nanopores/nanogaps have generated a lot of interest in ultrafast DNA sequencing. However, there are challenges to slow down the DNA translocation process to achieve a single-nucleobase resolution. A series of computational tools have been used in an attempt to study the DNA translocations in several model systems. The prospect of finding an efficient nanoelectrode for such human-genome sequencing might offer an entirely innovative way of preventive health care. Here, we have studied the performance of a boron-carbide (BC$_3$) based nanogap setup for DNA sequencing using the density functional theory and non-equilibrium Green's function-based methods. The electric current variations under different applied bias voltages are found to be significant due to changes in the nucleotides orientation and lateral position and can even outperform graphene. Computed relatively lower interaction energy for BC$_3$ electrodes compared to graphene electrodes indicates that BC$_3$ is a better nanoelectrode for DNA sequencing. From our results, we have found that the unique identification of all four nucleotides possible in the $0.3 - 0.4$ V bias region. Furthermore, each of the four nucleotides exhibits around one order of current difference, which makes it possible to identify all four nucleotides uniquely. Thus, we believe that BC$_3$ based nanoelectrodes may be utilized toward the development of a practical nanodevice for DNA sequencing.

## 1. Introduction

The development of personalized based medicine depends on individual genetic information, which needs techniques that can sequence DNA rapidly and inexpensively.[1,2] The Sanger's DNA sequencing method is very much capable of genome sequencing, but the practical usages are quite limited due to cost and sequence time. Several advanced techniques have been developed for this but the cost of sequencing is still far from the ideal price set by the "$1000-Genome" project.[3,4] Single-molecule sequencing technologies, especially nanopore-based DNA sequencing, can potentially sequence the long strands of DNA nucleobases without amplification or labels, but a standard set up yet to be established.[5-11]

Nanopores have been presaged as a promising way of providing low-cost and rapid DNA sequencing by measuring either ionic current as single-stranded DNA (ssDNA) translocated through a nanopore/nanogap or transverse electric currents across the nanopore/nanogap itself.[12-16,20,21] Furthermore, it allows sensing each nucleobase of ssDNA translocated through a nanopore. In particular, advanced nanopore/nanogap-based set up is studied for the application of DNA sequencing because it can determine the precise order of all four DNA nucleobases [adenine (A), guanine (G), thymine (T), and cytosine (C)].[15,16] Moreover, the nanopore/nanogap-based DNA sequencing technology does not require polymer chain reaction, amplification and high-cost chemical labeling of the sample. The nanopore technique can be of two types, such as biological nanopores and solid-state nanopore/nanogap. However, a solid-state nanomaterials based nanopore/nanogap approach is the most promising technology over biological nanopores to achieve a single-nucleobase resolution.[13,14] This is due to the higher resolution of 2D nanopores/nanogaps and high sensitivity of the solid-state nanopore/nanogap.[16-19] Therefore, nanopores/nanogaps in two-dimensional (2D) solid-state nanomaterials have attracted a lot of attention for the next-generation DNA sequencing.[20,21]

Especially, DNA sequencing using graphene (atomically-thick) sheet is very realistic and now many theoretical and experimental studies of DNA sequencing using graphene nanopore have also been proliferated.[19-23] In 2010, three different groups independently reported the fluctuations of ionic current when a double-stranded DNA (dsDNA) translocate through an atomically thick graphene-based nanopore.[13,14,24] Thereafter, in 2011, Venkatesan and co-workers have fabricated multi-layered graphene-$Al_2O_3$ based nanopore membranes for DNA and protein detection.[8] Although graphene-based nanopores/nanogaps are very promising for DNA sequencing, still it has some drawbacks such as high signal-to-noise ratios in nanopore experiments and geometry deformation after insertion of nucleobase in nanopore which hindered its practical use in DNA sequencing.[20,21,24,25] For example, it has also been reported that graphene edges are reactive in nature, due to favorable π-stacking interaction.[2,16,20,21,31] Hence, these issues have promoted researchers to study other atomically thick 2D monolayers for DNA sequencing.[26-34]

In this work, we aim to investigate an atomically thick boron-carbide ($BC_3$) nanogap setup for DNA sequencing. Zhang and co-workers have demonstrated that $BC_3$ monolayer is a dynamically stable material.[35] Atomically thick $BC_3$ with graphene-like structure has been synthesized, which shows similar properties like graphene.[36,37] The nearly close atomic size of boron (B) and carbon (C) atoms and their capability to form strong covalent (chemical) bonds, found to be important for many applications.[38,39] Despite several theoretical studies and experimental realizations of the BC$_3$ sheet, the available information concerning their intrinsic chemical and physical properties and DNA sequencing applications are yet to be explored. We believe that the polar nature of the BC$_3$ group compared to graphene may play an interesting role in terms of electrostatic interactions between the nanogap and nanoelectrodes. This may improve the coupling between BC$_3$ nanoelectrodes and the DNA nucleotides.

Therefore, it would be interesting to investigate whether the H-terminated BC$_3$ edges compared to H-terminated graphene edges can be helpful for controlled and rapid DNA sequencing or not. Motivated by all these reports, we have perceived the idea that such an atomically thick BC$_3$ monolayer can be a suitable nanodevice for DNA sequencing applications. Herein, we have used first-principles density functional theory (DFT) combined with non-equilibrium Green's function (NEGF) method[40-43] to study the structural and electronic transport properties of DNA nucleotides placed between the BC$_3$ nanoelectrodes. We have investigated the zero-bias transmission ($G_0$), transverse electric current ($I-V$ characteristics), and the bias-dependent transmission function for the four target nucleotides while translocated through the proposed nanogap setup. In experiments, the DNA molecule translocates/rotates while passing through the nanopore/nanogap. Therefore, the rotational and lateral translation effects account for all these in our modeling. It is, therefore, imperative to consider how the transmission function depends on the orientation of a given target nucleotide. Hence, we have also tried to cover all the possible orientation and lateral translational effects on the transmission function, which might be helpful during the experimental measurements.

## 2. Model and Computational Methods

The nanogap setup we focus on here is made up of atomically thick $BC_3$ nanoelectrodes. $BC_3$ is used as both nanoelectrodes, and central scattering region to avoid the contact resistance and to fulfill the condition of the sub-nanometer of the setup. The zigzag edges of $BC_3$ in the proposed nanogap setup are functionalized with hydrogen (H) atoms (**Figure 1**) since H terminated edges are easy to be realized experimentally.[40-43] Moreover, the $BC_3$ edge terminated with H forms the nanogap, which can facilitate the interaction between $BC_3$ electrode and target nucleotide. We have considered four DNA nucleotides (deoxyadenosine monophosphate (dAMP), deoxyguanosine monophosphate (dGMP), deoxythymidine monophosphate (dTMP), and deoxycytidine monophosphate (dCMP)) as our target nucleotides,[15,32,50-53] which are placed between the $BC_3$ nanoelectrodes as shown in **Figure 1**.

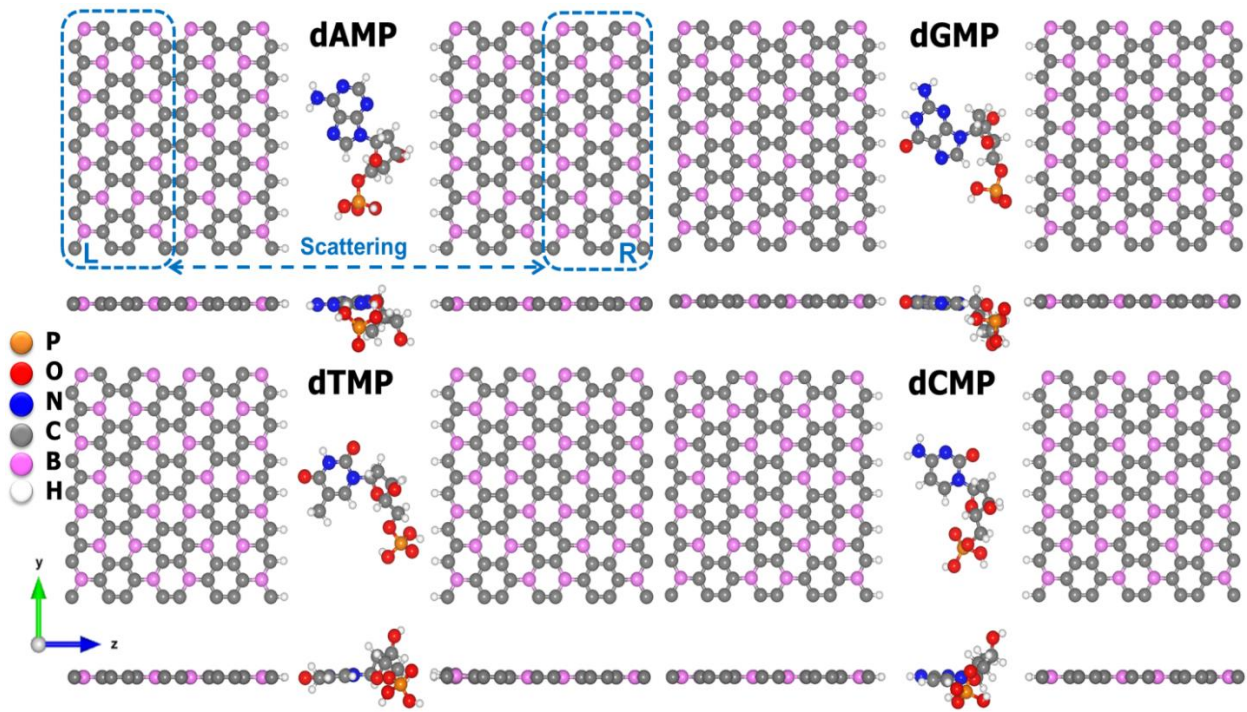

**Figure 1**. Top and side views of our proposed nanogap setup with the four target nucleotides (dAMP, dGMP, dTMP, and dCMP). Illustrating the left (L), right (R) electrodes and scattering

region. Here, orientations corresponding to $0°$ for all the nucleotides and $z$−axis is the transport direction.

Initially, we have optimized the isolated DNA nucleotides using the B3LYP/6-31G* level of theory as implemented in the Gaussian 09 code.[44] In the BC$_3$ device (**Figure 1**), a nanogap of ∼1.26 nm (H to H distance) is created along the armchair direction for the structural, electronic, and transport calculations. We have fully relaxed (all atoms) the BC$_3$ nanogap device with a target nucleotide inside the nanogap. The GGA-PBE method with double-zeta polarized (DZP) basis sets have been used for all the calculations as implemented in the SIESTA code.[45-47] Norm-conserved Troullier-Martins pseudopotentials are utilized to describe the interactions between the core and valence electrons.[48] We have used 200 Ry of mesh cut-off for real space integration, conjugate gradient algorithm for relaxation, and $1 \times 5 \times 3$ k-grid for Brillouin-zone sampling. The convergence tolerance in density matrix is set to $10^{-4}$ eV and the residual forces criteria on the atoms are set to $< 10^{-2}$ eV/Å.

We have also considered rotation effects for structural optimization, zero-bias transmission, density of states, and I − V characteristics. For this, we have rotated each target nucleotides in the yz−plane around x−axis from 0° to 180° in the steps of 30° as shown in **Scheme 1 [Figure S1** and **S2**; **Supporting Information (SI)]**. The whole nanogap setup (electrode+nucleotide) is then fully relaxed (at 30°, 60°, 90°, 120°, 150°, and 180°) to achieve the energetically most stable configuration. The relative energy of these rotated configurations is also tabulated in **Table S1**. We have found that dAMP and dCMP are energetically most stable with $90°$ orientation, while 180° orientation is the most stable configuration for dGMP and dTMP, as shown in **Figure S3**.

Furthermore, the effects of nanogap width, lateral translation, and out-of-plane on the transverse transmission have been investigated for the most stable system for each nucleotide. For nanogap width, we have increased the nanogap size from 1.26 nm to 1.47 nm for all the four DNA nucleotides as shown in **Scheme 2** [**Figure S4**; **SI**]. The whole setup (electrode+nucleotide) is then fully relaxed for each target nucleotide to compute the transverse transmission at zero-bias. We have also laterally translated each target nucleotide inside the nanogap along the $z-$axis in the positive and negative direction by $\pm 0.5$ Å, as shown in **Scheme 3** [**Figure S5**; **SI**]. Furthermore, we have moved DNA nucleotides out-of-plane along the $x-$axis in the $yz-$plane for all four nucleotides, as shown in **Scheme 4** [**Figure S6**; **SI**].

The following equation was used to calculate the interaction energy ($E_i$) of the grapheme-poly-G and $BC_3$-poly-G systems:

$$E_i = E_{graphene/BC_3+poly-G} - (E_{poly-G} + E_{graphene/BC_3}) \qquad (1)$$

where $E_{graphene/BC_3+poly-G}$, $E_{poly-G}$ and $E_{graphene/BC_3}$ are the total energy of the fully relaxed graphene/$BC_3$ device with target poly-G nucleotide molecule, isolated poly-G, and graphene/$BC_3$. Both the GGA-PBE[45-57] and vdW-DF-CX functionals[52] are used to calculate the interaction energies as implemented in the SIESTA code.

Then the transverse transmission and I-V characteristics have been calculated for the fully relaxed device (electrode+nucleotide). The transport calculations are performed by utilizing the Landauer-Buttiker formalism.[49] For this, we have used the NEGF approach combined with DFT (DFT-NEGF), as implemented in the TranSIESTA code.[45] The basis sets and real-space integration are the same as those described above for the geometrical relaxation part. For such calculations, the proposed setup (Figure 1) defined into three parts: (i) $BC_3$ right electrode (R),

(ii) central scattering (device region), (iii) BC$_3$ left electrode (L). The electric current (I) is calculated from the integration of the transmission function as follows:

$$I(V_b) = \frac{2e}{h} \int_{\mu_R}^{\mu_L} T(E, V_b) \left[ f(E - \mu_L) - f(E - \mu_R) \right] dE \quad (2)$$

where h is Planck's constant, e is the electron charge, $T(E, V_b)$ is the transmission of the setup and $f(E - \mu_L)$ and $f(E - \mu_R)$ are the Fermi functions for the electrons in the left (L) and right (R) nanoelectrodes, respectively. The transmission $T(E, V_b)$ at the L and R nanoelectrodes depends on the nature of contact between the BC$_3$ nanoelectrodes and the target DNA nucleotide and consequently, the current in equation 2 follows the same trends.[45,49-51,53]

## 3. Results and Discussion

We have first compared the interaction energy of DNA nucleobases towards the BC$_3$ and graphene electrode edges. This is essential since, in a practical DNA sequencing setup, fast detection of the DNA nucleobases is very much desirable. If the nanogap/nanopore edges tend to interact strongly with the translocating nucleotide molecule, it might not be an ideal nanoelectrode device. Since graphene nanopores very quickly exhibit clogging of the pore, leading to blockage. This can be attributed to strong interactions between translocating DNA nucleotides and graphene nanopores/nanogaps, yielding sticking and irreversible pore closure.[2,20,54] Therefore herein, a comparative study of the interaction of poly-guanine (poly-G; dimer) with BC$_3$ and graphene edge has been discussed. The fully optimized atomic structures of both the systems are shown in **Figure S7**. We have found that the interaction energy values increase significantly when vdW corrections included. For example, with vdW correction, the calculated interaction energy value for graphene-poly-G is increased to -2.69 eV from -1.63 eV (without vdW correction). Similarly, the respective values are -2.24 eV (with vdW correction) and −1.13 eV (without vdW correction) for BC$_3$-poly-G system. So, the vdW corrected interaction energies show that graphene interacts strongly compared to BC$_3$. The higher interaction energy for graphene electrode edges confirms the strong interaction nature of the graphene electrode edges, which may block the pore. Thus, the BC$_3$ electrode indicates that this can be a comparatively better nanoelectrode in a nanogap device for DNA detection.

In the context of our proposed BC$_3$ nanogap setup functioning, the orientations of the DNA nucleotides concerning the BC$_3$ nanoelectrodes play an essential role. Because during the translocation process of an ssDNA through the BC$_3$ nanogap, several different orientations of all four nucleotides are possible concerning BC$_3$ nanoelectrodes. It is, therefore, essential to consider the orientation effects on the transmission function of a given nucleotide. Consequently, we have considered the rotations of all four target nucleotides around $x$-axis in the $yz$ plane in the steps of 30° [**Scheme 1 (Figure S1** and **S2**); SI]. The relative energy of these rotated configurations with respect to the 0° (original) orientation is also tabulated in **Table S1**. We have found that dAMP and dCMP are energetically most stable with 90° orientation, while 180° orientation is the most stable configuration for dGMP and dTMP (**Figure S1-S3; SI**). **Figure S3** shows that at 180° orientation, the dGMP and dTMP nucleotide molecules are perfectly aligned for strong H-bonding interactions with the hydrogen terminated BC$_3$ electrode. The same has been noticed for dAMP and dCMP at 90° orientation.

We have calculated zero-bias transmission ($G_0$) functions for the energetically most stable configuration of each target nucleotides to understand the nucleotide-electrode coupling as a function of energy (**Figure 2**). We find that distinct molecular states appear in transmission function for all the four DNA nucleotides in the ±1.0 energy window. To understand further, we have investigated the zero-bias density of states of the BC$_3$ nanogap device with a target nucleotides. Our zero-bias DOS (**Figure S8**) is in well corresponds to the transmission peaks. Further, to understand transmission peaks around the Fermi region (1.0 to -1.0 eV energy window), we plotted the eigenchannels (molecular orbitals) calculated using the Inelastica package. In the negative energy region (0 to −1.0 eV), HOMO peaks (**Figure 2b**) have been identified for all the four nucleotides, which corresponds to the sharp transmission peaks that appear around the Fermi energy. The relative position of the sharp transmission peaks can give us an idea about the extent of coupling between the electrode and nucleotides, which can be very useful for their distinction. However, we have not found any LUMO peak close to the Fermi energy. **Figure 2c** presents the HOMOs that correspond to sharp transmission peaks that appear around the Fermi-level. The negative and positive lobes are shown in red and green colours, respectively. The HOMO energies are shown to show the peak position with respect to the Fermi energy. The HOMOs corresponding to the transmission peaks just below the Fermi-level are found to be localized on the purine and pyrimidine nucleobases (**Figure 2c**). Thus, the zero-bias transmission function depends on the types of nucleotides. This could be because of two different groups of nucleotides, i.e., pyrimidine nucleobases [thymine (T) and cytosine (C)] and purine nucleobases [adenine (A) and guanine (G)]. The main characteristics of physical property between these two groups are the nucleobase size i.e., purine nucleobases are more significant than the pyrimidine nucleobases. This could be the reason for smaller separation and robust coupling of dAMP and dGMP nucleotides with the $BC_3$ electrodes. Comparing our obtained results of dAMP and dGMP, we have found that the Fermi energy is associated closely with the HOMO peak of dGMP at $E(G) = -0.10$ eV, while for dAMP seems at $E(A) = -0.13$ eV. For dCMP and dTMP, we have found that the Fermi energy is associated closely with the HOMO peak of dCMP at $E(C) = -0.21$ eV, whereas for dTMP seems at $E(T) = -0.71$ eV. This result infers us that all four nucleotides may give a noteworthy difference in the conductivity at low applied bias voltages.[15]

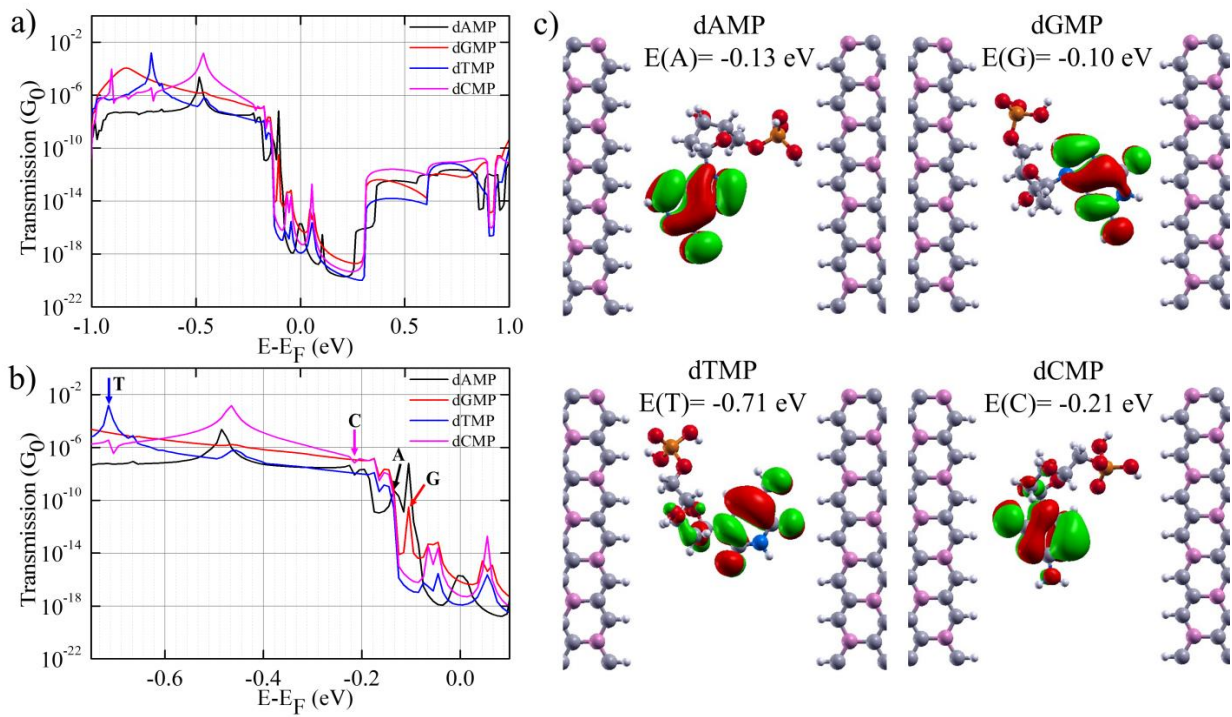

**Figure 2**. (a) Zero-bias transmission function curve plotted on a logarithmic scale for all four different nucleotides (dAMP, dGMP, dTMP, and dCMP). (b) Zoomed zero-bias transmission function plot to identify molecular orbitals responsible for transmission peaks for different nucleotides. (c) Isosurface plots of the highest occupied molecular orbitals (HOMOs) responsible for the sharp transmission peaks. The Fermi level is aligned to zero.

Furthermore, the transmission values and the position of the resonance peak are related to the nucleotide orientation. When a DNA nucleotide is rotated, the resonance peak position shifts downward/upward with respect to the Fermi energy. Consequently, the magnitude of the transmission changes as presented in **Figure S8**. This could be because of the electrode nucleotide coupling change. However, the pattern of the transmission function remains the same and we have not observed any specific trend for this. This is understandable as the extent of coupling can be different due to orientations and alignments of the nucleotide. We wanted to see if the transmission value changes significantly due to rotation. If it does not change much, then it may be easier to detect. Further, we note that for all orientations considered here, the transmission function $T(-1 \text{ eV} \leq E \leq 1$ eV, at bias (V=0) of purine nucleobases (dAMP and dGMP) ranges from $10^{-21}$ to $10^{-3}$ $G_0$, while in the case of pyrimidine nucleobases (dTMP and dCMP) the corresponding transmission function can range from $10^{-22}$ to $10^{-2}$ $G_0$ (**Figure S9**). Furthermore, when we compare our transmission spectra due to rotation of the nucleotide (at 30°, 60°, 90°, 120°, 150°, and 180°), we find smaller variations in the transmission function compared to the transmission spectra of graphene and graphene-hBN.[15,31] This can be promising for the detection of nucleotides.

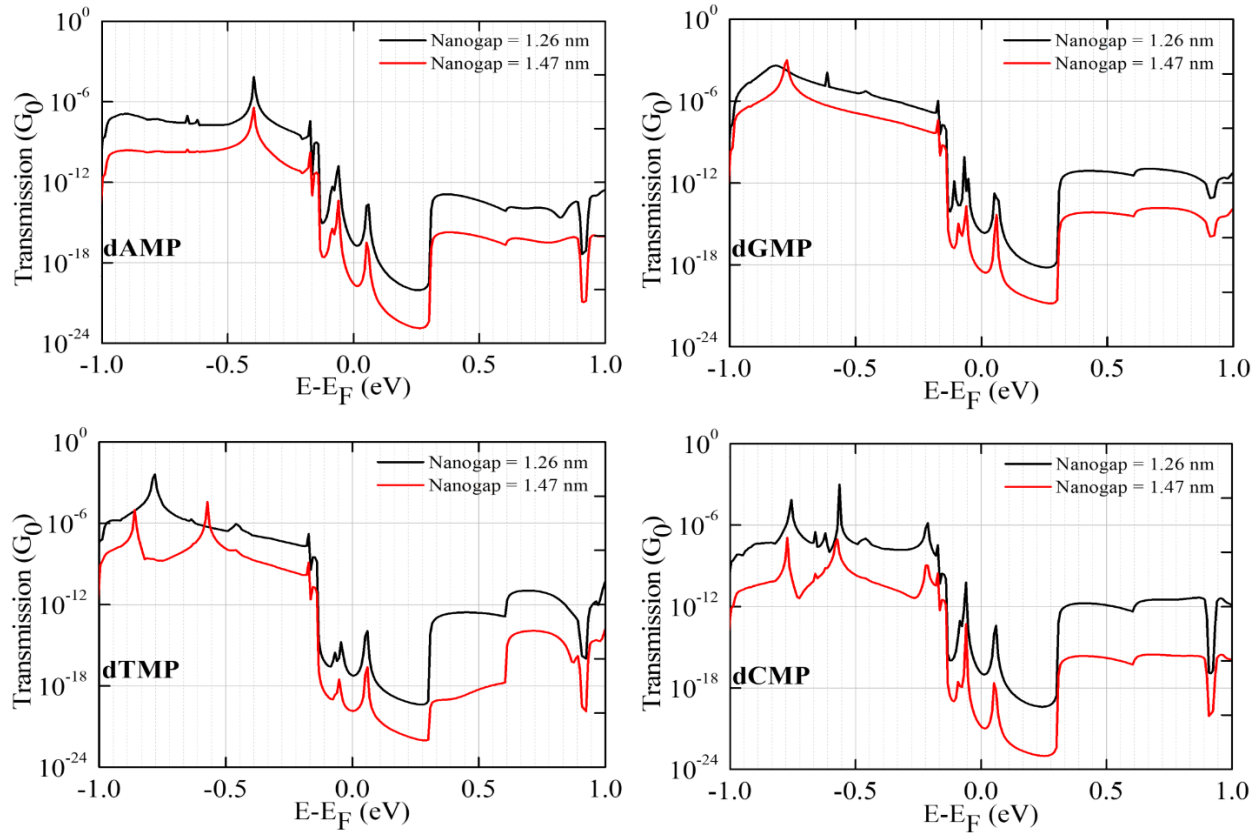

**Figure 3**. The effect of the nanogap width on the zero-bias transmission as a function of energy for the four nucleotides, dAMP, dGMP, dTMP, and dCMP.

Additionally, to realize how the transmission function is affected not only by nucleotide rotation but also by nanogap width and lateral translation of a given DNA nucleotide, we have investigated the nanogap width effect on the transmission function. We find that the transmission spectra drop exponentially when the distance between the nucleotide and electrode increases.[15] Herein, one can see that when the nanogap width is increased from 1.26 nm to 1.47 nm, the transmission drops significantly for all the four nucleotides (**Figure 3**). A shift of HOMO peak position is seen for the dTMP case. Though, for similar orientations of the DNA nucleotides within such a wider gap, the electrode-nucleotide molecule overlap depends on the localization of the HOMO peaks. This allows improved differentiation between the purine and pyrimidine types nucleobase.

Furthermore, lateral translation of the nucleotides in the nanogap causes a shift in the resonance peak position within the $yz-$plane. **Figure 4** represents the resulting change in the transmission spectra with lateral translations of each nucleotide in steps of $\pm 0.5$ Å along the $z-$axis. We note that for all four nucleotides, the transmission peak shifts and peak width are found to be the same. A shift of the HOMO peak position is seen for rotation of all four nucleotides relative to the $BC_3$ edges. The behaviors of the shifts and the peak widths are found to be qualitatively the same for all four nucleotides. The corresponding DOS has also confirmed this finding (**Figure S10**). Similarly, another strong effect comes from the out-of-plane fluctuation of the nucleotide. To understand this, we have computed the zero-bias transmission function as shown in **Figure S11**. We find that the transmission pattern remains nearly the same; however due to the weakening of electrode-nucleotide coupling, the peak position shifts with respect to the Fermi level. Therefore, we have not observed a much stronger effect due to out-of-plane fluctuation. Therefore, the transmission function plots indicate that the differences in chemical properties and physical structures between the pyrimidine nucleobases (dCMP, dTMP) and purine nucleobases (dAMP, dGMP) certainly affect the coupling strength of nucleobases with the $BC_3$ electrode. This is, in turn, promising for distinguishing these nucleobases through the $BC_3$ electrode.

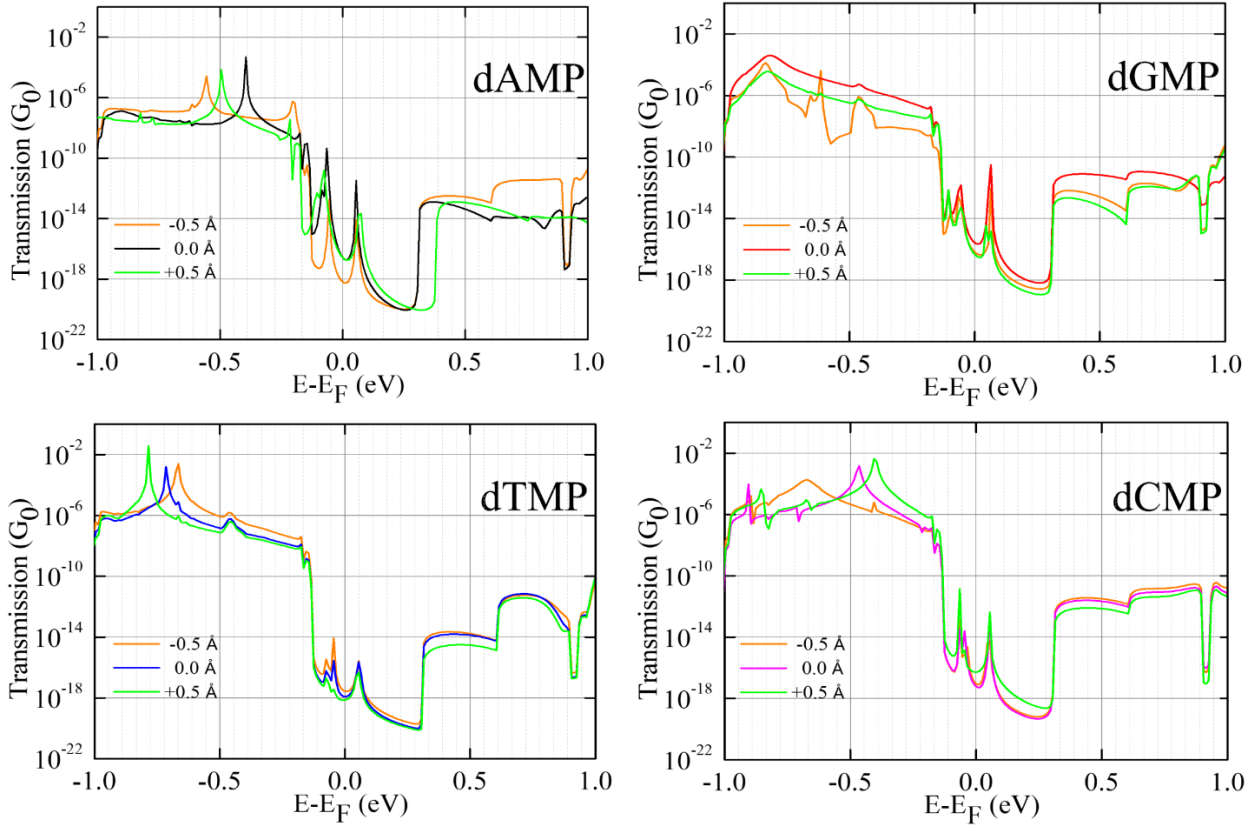

**Figure 4**. The change in the zero-bias transmission function due to the lateral translation in steps of $\pm 0.5$ Å along the $z$-axis for all the four nucleotides (dAMP, dGMP, dTMP, and dCMP).

Next, we have investigated the $I-V$ characteristics of the most promising properties for DNA nucleotides detection. **Figure 5** shows the $I-V$ patterns for the energetically most stable configurations of the four target nucleotides. The identification of these four nucleotides looks possible within the $0.3 - 0.4$ V bias window. At 0.4 V, different current signals observed for all four nucleotides in the following order: $I_{dCMP} > I_{dGMP} > I_{dAMP} > I_{dTMP}$. **Figure 5b** shows the read-outs of current signals of the four target nucleotides within a given bias window ($0.3 - 0.4$ V). This figure also shows that the four nucleotides transmit unique current signals at the given bias voltage window. More clearly, dCMP and dTMP give higher and lower current respectively, whereas dGMP and dAMP can be identified in between the former two nucleotides. Hence, we find robust sensitivity in the given bias window. Our calculated current values are higher than $10^{-8}$ nA. Makusu et al. have shown that such current (in the nA range) values can be possible to measure experimentally.[59] Prasongkit and co-workers have also stated that if the current values are below $10^{-11}$ nA, then they are expected to disappear into the electrical background noise in experimental measurements.[15] Therefore, our calculated current values are possible to measure and can be used to detect the nucleotides. It is worth mentioning that the bias in our proposed nanogap setup is relatively low compared to graphene[15,51] and graphene-hBN[31], and phosphorene[32] for DNA sequencing. Further, one would expect an increase in the current with bias. However, that does not happen due to other factors such as negative differential resistance, coupling, and so on. For example, we have four different nucleobases, and the orbital interactions (coupling) not necessarily be in the same order of their interaction energies or their sizes. To understand further, we have analysed the bias dependent transmission spectra (**Figure 6**). At zero-bias transmission (**Figure 2**), the HOMO peak ($-0.10$ eV) of dGMP appears close to the Fermi level, while the HOMO peak is at $-0.13$ eV for dAMP. For dCMP, the HOMO peak is at $-0.21$ eV, and at $-0.71$ eV for dTMP. Hence, at 0.1 V bias, we get a better current for dGMP and dAMP. However, with an increasing bias voltage, the trend changes. So, this has to do with the coupling between the electrode and nucleotide orbitals. The change in the position of the HOMO/LUMO peaks with the change in bias is also very important. That means molecular states present in the system will play an important role. [15,31,51,58] **Figure 6** reveals the molecular states are responsible for the transmission, and the change in transmission function depends on the applied bias voltage. The shift in the transmission resonance peaks is observed towards the Fermi energy as the applied bias voltage increases. However, in the case of dCMP, we find that the HOMO peak exhibits significantly maximum movement and thus adds up to the transmission spectra. This is revealed in the $I-V$ curve as dCMP shows the highest current among all four target nucleotides. Moreover, the HOMO peak movement is rather slow in the case of dGMP, dAMP, and dTMP with changing bias. Consequently, the current is significantly lower for all three nucleotides. This lead crossing in the $I-V$ signals is happening. This is in very much in agreement with our $I-V$ characteristics curve too.

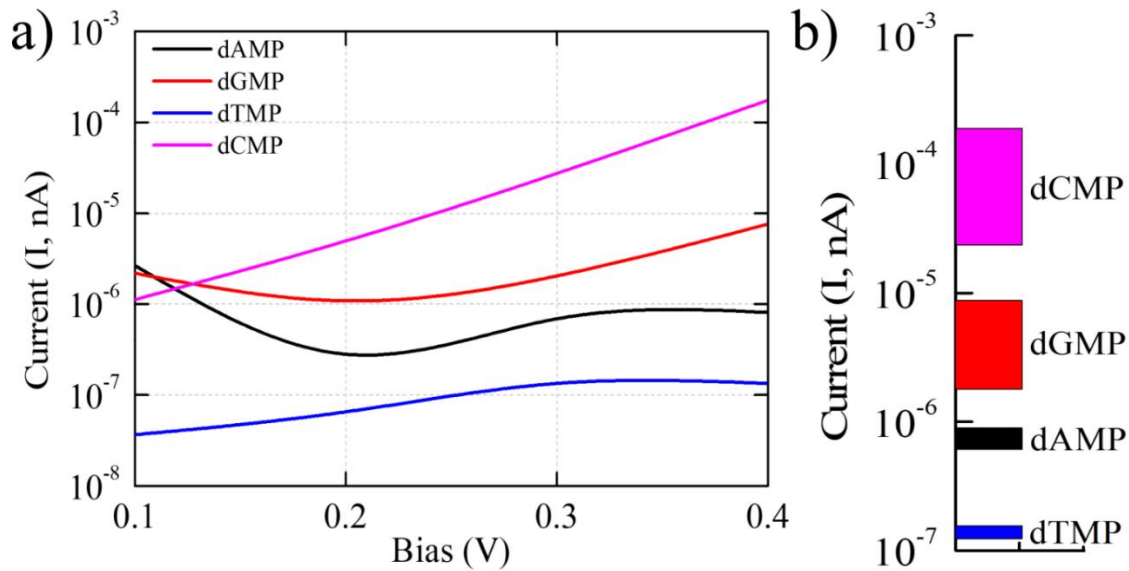

**Figure 5**. (a) $I-V$ Characteristics curve plotted on a logarithmic scale of all four nucleotides (dAMP, dGMP, dTMP, and dCMP) in the $BC_3$ nanogap setup. (b) The current responses (read-outs of current traces from the $I-V$ curve) for all four nucleotides within the bias range of 0.3-0.4 V.

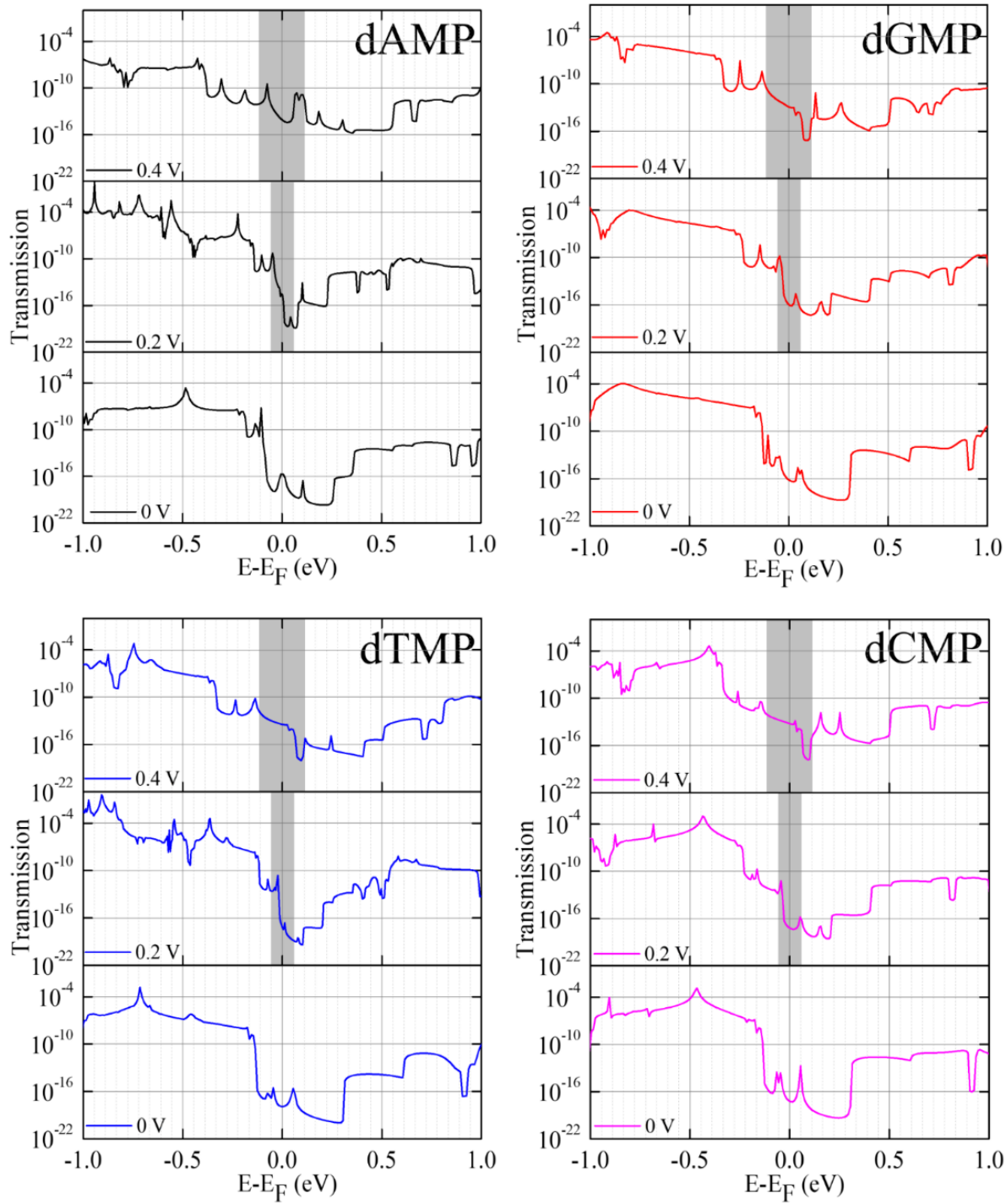

**Figure 6**. The bias-dependent transmission function of all the four nucleotides, dAMP (top left panel), dGMP (top right panel), dTMP (bottom left panel), and dCMP (bottom right panel), with the variation of energy E. The Fermi level is set to zero. The grey shaded area represents the bias voltage window of $\pm V/2$.

We have made additional efforts to calculate the $I-V$ curves when the DNA nucleotides are rotated in the nanogap (from $0°$ to $180°$ in the steps of $30°$) in the manner discussed above. This is important because, in a practical DNA sequencing nanogap setup, the current may vary by several orders of magnitude. **Figure S12** represents the possible range of current values for a bias of 1 V for all four nucleotides at different orientations. At 0.3 V bias voltage (**Figure S13**) the current variation is noted in the following order: $I_{dAMP} > I_{dGMP} > I_{dTMP} > I_{dCMP}$. We find that there is a substantial overlap when considering all rotations is observed. Therefore, it may be difficult to distinguish all the four nucleotides (**Figure S13**).

In our study, we neither included the solvent effect nor incorporated molecular dynamics simulation (MDS), which could lead to different molecular orientations inside the proposed nanogap setup.[53] Indeed, Feliciano et al., have demonstrated the effectiveness of the solvent (water) medium for the zero-bias transmission of the graphene-nucleobase system.[55] The small changes observed in the transmission functions indeed describe the effect of screening, assisted by the water molecules. However, this will only affect the transverse electric current magnitude. A solvent medium could deliver better or less transverse electrical current signals. But the trend observed within systems doesn't change. In a similar report by Lagerqvist et al., have shown the structural fluctuations in the DNA nucleotides causing the change in transmission function.[56,57] However, they have concluded that the direct effect of water on transmission function can be neglected. In conclusion, one can say that the separation of nucleotides based on transverse electric current (with or without solvent medium) is entirely reasonable as the trend will not be changed. However, there can be a change in the transverse electrical current magnitude. Hence, investigating the effects of the medium opens the doors for future studies, which could be accomplished using a hybrid QM/MM method.[53,55]

## 4. Conclusions

In conclusion, we have demonstrated the applicability of the $BC_3$ based nanogap setup for DNA sequencing. We find that the $BC_3$ zigzag edge terminated with H forms the nanogap, which significantly improves the interaction between the $BC_3$ electrodes and target nucleotide. The molecular orbitals corresponding to the transmission peaks just below the Fermi energy are found to be localized on the purine (A and G) and pyrimidine (C and T) nucleobases. Consequently, the transmission function depends on the chemical and physical properties of DNA nucleobase. The transmission values and the position of the transmission resonance peak are related to the nucleotide orientation; as a result, the magnitude of the transmission also changes. Moreover, increased nanogap width shows that transmission drops significantly for the target nucleotide. Lateral translation of the nucleotides in the nanogap causes a shift in the resonance peak position; however, peak width is found to be the same. The analysis of $I - V$ characteristic curves for specific orientations of all four nucleotides reveals that the $BC_3$ based nanogap sequencing setup is indeed feasible for the unique identification of the target nucleotide. We have found that the unique identification of all the four nucleotides could be possible within the $0.3 - 0.4$ V bias region. We have noted that dCMP and dTMP give the highest and lowest current signals, whereas dGMP and dAMP can be identified in between the former two nucleotides. Thus, each of the four nucleotides exhibits around one order of current difference, which makes possible unique identification of these nucleotides. Furthermore, relatively lower interaction energy for $BC_3$ electrodes compared to graphene electrodes indicates that $BC_3$ may be a better nanoelectrode for DNA sequencing. Thus, based on our analysis, we believe that the $BC_3$ nanogap based device may be promising for controlled and rapid human DNA sequencing.

## 5. Associated Contents

* Supporting Information

## 6. Conflicts of interest

There are no conflicts of interest to declare.

## 7. Acknowledgments

We thank IIT Indore for the lab and computing facilities. This work is supported by DST-SERB, (Project Number: EMR/2015/002057) New Delhi and CSIR [Grant number: 01(2886)/17/EMR (II)] and (Project Number: CRG/2018/001131) and SPARC/2018-2019/P116/SL. R. L. K., P. G. and G. B. thanks MHRD for research fellowships. We would like to thank Dr. Vivekanand Shukla for fruitful discussion throughout this work.

## 8. ORCID

Rameshwar L. Kumawat: 0000-0002-2210-3428

Biswarup Pathak: 0000-0002-9972-9947

**Table of Content (TOC):**

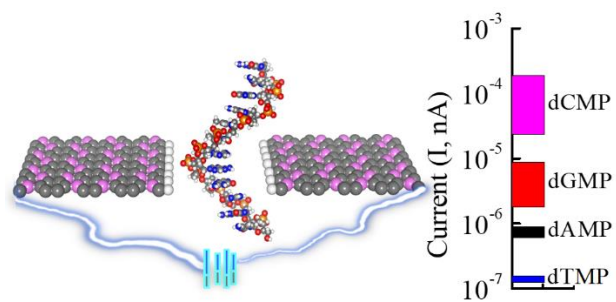